\documentclass[preprint2]{aastex6}
\usepackage{multirow}

\begin{document}

\title{The Core Mass Function Across Galactic Environments. II.\\Infrared Dark Cloud Clumps}

\author{Mengyao Liu\altaffilmark{1}, Jonathan C. Tan\altaffilmark{1,2}, Yu Cheng\altaffilmark{3}, Shuo Kong\altaffilmark{4}}
\affil{
$^1$Dept. of Astronomy, University of Virginia, Charlottesville, Virginia 22904, USA\\
$^2$Dept. of Space, Earth \& Environment, Chalmers University of Technology, Gothenburg, Sweden\\
$^3$Dept. of Astronomy, University of Florida, Gainesville, Florida 32611, USA\\
$^4$Dept. of Astronomy, Yale University, New Haven, CT 06511, USA}





\begin{abstract}
We study the core mass function (CMF) within 32 dense clumps in seven
infrared dark clouds (IRDCs) with the {\it Atacama Large
Millimeter/submillimeter Array} ({\it ALMA}) via 1.3~mm continuum
emission at a resolution of $\sim$1\arcsec. We have identified 107
cores with the dendrogram algorithm, with a median radius of about
0.02~pc. Their masses range from 0.261 to 178~$M_{\odot}$. 
After applying completeness corrections, we fit the combined IRDC CMF with a
power law of the form $d N / d\:{\rm log} M \propto M^{-\alpha}$ and
derive an index of $\alpha\simeq0.86\pm0.11$ for $M \geq
0.79\:M_\odot$ and $\alpha\simeq0.70\pm0.13$ for $M\geq
1.26\:M_\odot$, which is a significantly more top-heavy distribution
than the Salpeter stellar initial mass function (IMF) index of
1.35. We also make a direct comparison of these IRDC clump CMF results to those measured
in the more evolved protocluster G286 derived with similar methods,
which have $\alpha\simeq1.29\pm0.19$ and $1.08\pm0.27$ in these mass
ranges, respectively. These results provide a hint that, especially
for the $M\geq 1.26\:M_\odot$ range where completeness corrections are
modest, the CMF in high pressure, early-stage environments of IRDC
clumps may be top-heavy compared to that in the more evolved, global
environment of the G286 protoclusters. However, larger samples of
cores probing these different environments are needed to better
establish the robustness of this potential CMF variation.
\end{abstract}

\keywords{stars: formation -- ISM: clouds}

\section{Introduction} 

The origin of the stellar initial mass function (IMF) remains one of
the most important unsolved problems in astrophysics. In general, the
IMF can be described as having a broad peak just below $1\:M_\odot$,
similar in shape to a log normal, but then extending with a power law form at
high masses (see, e.g., Bastian et al. 2010), i.e.,
\begin{equation}  
\frac{{\rm d} N}{{\rm d} {\rm log} M}\propto{M}^{-\alpha}.
\label{eq:pl}
\end{equation}
Salpeter (1955) derived $\alpha\simeq1.35$ between 0.4 and
10~$M_{\odot}$ and this value has remained valid as the standard
description of the $\gtrsim 1\:M_\odot$ IMF from more recent studies.

Observations of dense cores show that the core mass function (CMF) may
be similar in shape to the IMF (e.g., Alves et al. 2007; Andr{\'e} et
al. 2010; Offner et al. 2014; K\"onyves et al. 2015; Ohashi et
al. 2016; Cheng et al. 2018). Such a similarity is taken as evidence
that the stellar IMF is in large part determined by the fragmentation
process in molecular clouds, after also allowing for a core to star
formation efficiency. However, to most accurately test such a scenario, then
observationally one should ideally measure the pre-stellar core (PSC)
mass function, with PSCs being cores at an evolutionary stage just
before the onset of star formation. This method has been carried
  out using FIR {\it Herschel} imaging of nearby regions, such as
  Aquila ($d=$260~pc), by, e.g., K\"onyves et al. (2015), who find a
  pre-stellar core mass function (PSCMF) that is similar in shape to
  the stellar IMF.

Unfortunately identifying PSCs in more distant star-forming regions is
a non-trivial task. Using mm continuum emission to identify cores,
i.e., the thermal emission from dust, is the typical method adopted
(and will be the one used in this paper). This then allows a measure
of the mass of the sources, assuming given dust emissivities,
dust-to-gas mass ratio and dust temperature. At this point the sample
likely contains a mixture of prestellar cores and protostellar cores,
and with the latter tending to be more easily detected given their
internal heating.  Attempts can then be made to remove obvious
protostellar sources, e.g., those cores associated with infrared or
x-ray emission or with outflow tracers. Such an approach was adopted
by Ohashi et al. (2016), who first identified 48 cores in IRDC
G14.225-0.506 from 3~mm continuum emission and then proposed 28 of
these to be PSCs, based on a lack of IR or x-ray emission.
However, in high column density regions
such as IRDCs, lack of detected IR emission, e.g., from {\it Spitzer}
MIPSGAL 24~$\rm \mu m$ images (Carey et al. 2009), is no guarantee a
core is pre-stellar, as found by, e.g., Tan et al. (2016), who find
that the presence of protostellar outflows, e.g., as traced by CO, can
be a more powerful probe of protostellar activity depending on the
extinction in the region. Furthermore, even if a core is identified as
being pre-stellar from the above methods, it is not clear at which
evolutionary stage it is at, i.e., whether it will grow much more in
mass before forming a star.

An alternative approach is to try and select PSCs that are on the
verge of forming stars via certain chemical species, especially
deuterated species, such as $\rm N_2D^+$ (see, e.g., Caselli \&
Ceccarelli 2012; Tan et al. 2013; Kong et al. 2017). However, this
requires very sensitive observations, and then the question of
measuring the masses of the PSCs still needs to be addressed, e.g.,
via associated mm continuum emission or dynamically via line widths
from some measured size scale.

Given the above challenges, a first step for distant regions is 
to characterize the combined pre-stellar and protostellar CMF, by
simply treating all the detected sources as cores of interest. This
approach has been adopted by, e.g., Beuther \& Schilke (2004), Zhang
et al. (2015), Cheng et al. (2018) and Motte et al. (2018).
Such an approach, which is the one we
will also adopt in this paper, is really a measurement of the mm luminosity function
of ``cores'' with potentially a mixture of PSCs and protostellar cores
being included in the sample, although, it is the latter, being warmer,
that will tend to be identified in a given protocluster.

Since there are large potential systematic uncertainties associated
with both core identification and core mass measurement, it is
important to attempt to provide uniform and consistent observational
metrics of core populations in different star-forming regions and
environments to allow comparison of relative properties. With this
goal in mind, we derive the mm-continuum-based CMF from observations
of dense regions of Infrared Dark Clouds (IRDCs), thought to be
representative of early stages of massive star and star cluster
formation (see, e.g., Tan et al. 2014). Most importantly, we use the same methods as our
previous study of the more evolved protocluster G286.21+0.17
(hereafter G286) (Cheng et al. 2018, hereafter Paper I).

There have been several previous studies of clump and core mass
functions in IRDCs. Rathborne et al. (2006) measured an IRDC clump
($\sim$0.3 pc-scale) mass function, with high-end power law slope
$\alpha \simeq 1.1 \pm 0.4$ above a mass of $100\:M_{\odot}$ via
1.2~mm continuum emission. Ragan et al. (2009) identified structures
on $\sim$0.1pc scales and found $\alpha \simeq 0.76 \pm 0.05$ from 30
to 3000 $M_{\odot}$ through dust extinction.
Zhang et al. (2015) measured the masses of 38 dense cores (with
$\sim$0.01 pc scales) in the massive IRDC G28.34+0.06, clump P1 (also
known as C2 in the sample of Butler \& Tan 2009, 2012) via 1.3~mm
continuum emission and found a lack of cores in the range 1 to
2~$M_{\odot}$ compared with that expected from an extrapolation of the
observed higher-mass population with a Salpeter power law mass
function. Finally, as mentioned above, Ohashi et al. (2016) studied
IRDC G14.225-0.506 and identified 28 starless cores on scales
$\sim$0.03 pc and derived $\alpha \simeq 1.6 \pm 0.7$ from with masses
ranging from 2.4 to 14 $M_{\odot}$ via 3 mm dust continuum emission.

We have conducted a 1.3~mm continuum and line survey of 32 IRDC clumps
with {\it ALMA} in Cycle 2. These regions are of high mass
surface density, being selected from mid-infrared
(\textit{Spitzer}-IRAC 8 $\mu$m) extinction (MIREX) maps of 10 IRDCs
(A-J) (Butler \& Tan 2012). The distances to the sources, based on
near kinematic distance estimates, range from 2.4~kpc to 5.7~kpc. The
first goal of this survey was to identify PSCs via $\rm
N_{2}D^{+}(3-2)$ emission, with about 100 such core candidates
detected (Kong et al. 2017). Here we report on the analysis of the
1.3~mm continuum cores and derivation of the CMF in these 32 IRDC
clumps. In \S\ref{S:methods} we describe the observations and analysis
methods. In \S\ref{S:results} we present our results on the
construction of the CMF, including with completeness corrections, and
the comparison to G286. We discuss the implications of our results and
conclude in \S\ref{S:conclusions}.

\section{Observations and Analysis Methods}\label{S:methods}

\subsection{Observational Data}\label{S:obs}

We use data from ALMA Cycle 2 project 2013.1.00806.S (PI: Tan), which
observed 32 IRDC clumps on 04-Jan-2015, 10-Apr-2015 and 23-Apr-2015,
using 29 12 m antennas in the array. The total observation time
including calibration is 2.4~hr. The actual on-source time is
$\sim$2-3~min for each pointing (30 pointings in total).
 
The spectral set-up included a continuum band centered at 231.55~GHz
(LSRK frame) with width 1.875~GHz from 230.615~GHz to 232.490~GHz.  At
1.3~mm, the primary beam of the ALMA 12~m antennas is $27\arcsec$
(FWHM) and the largest recoverable scale for the array is $\sim
11\arcsec$ ($\sim$ 0.3 pc at a typical distance of 5 kpc). No ACA
observations were performed. The sample of 32 targets was divided into
two tracks, each containing 15 pointings.  Track 1, with reference
velocity of +58 $\rm km s^{-1}$, includes A1, A2, A3, B1, B2, C2, C3,
C4, C5, C6, C7, C8, C9, E1, E2 (following the nomenclature of Butler
\& Tan 2012).  Track 2, with reference velocity of +66 $\rm km
s^{-1}$, includes D1, D2 (also contains D4), D3, D5 (also contains
D7), D6, D8, D9, F3, F4, H1, H2, H3, H4, H5, H6.  The continuum image
reaches a 1$\sigma$ rms noise of $\sim$0.2~mJy in a synthesized beam
of $\sim$ 1.36$\arcsec$ $\times$ 0.82$\arcsec$. Other basebands were
tuned to observe N$_2$D$^{+}$(3-2), SiO(5-4), $\rm C^{18}O$(2-1),
DCN(3-2), DCO$^{+}$(3-2) and $\rm CH_{3}OH \ (5(1,4)-4(2,2))$. These
data have mostly been presented by Kong et al. (2017), with the
SiO(5-4) data to be presented by Liu et al. (in prep.).



To investigate the flux recovery of our 12m data, we use the
archival data from the Bolocam Galactic Plane Survey (BGPS) (Aguirre
et al. 2011; Ginsburg et al. 2013), which are the closest in
frequency single-dish millimeter data available. We measure the flux density in both ALMA and BGPS images 
of each clump (the aperture is 27$\arcsec$ across, i.e., one ALMA primary beam size)
and then convert the BGPS flux
density measurements at 267.8 GHz to the mean ALMA frequency of
231.6 GHz via $S_{\nu} \propto \nu ^{\alpha_\nu}$ assuming $\alpha_\nu = 3.5 \pm 0.5$. For the ALMA data we measure the total flux above a 3$\sigma$ noise level threshold. 
Finally we derive a median flux recovery
fraction of 0.19 $\pm$ 0.02. As expected, these 12m array only ALMA observations filter out most of the total continuum flux from the clumps.


\subsection{Core Identification}\label{S:methods_ci}

Our main objective is to identify cores using standard, reproducable
methods. In particular, we aim to follow the methods used
in our Paper I
study of the G286 protocluster as closely as
possible so that a direct comparison of the CMFs can be made. Thus for
our fiducial core finding algorithm we will adopt the dendrogram
(Rosolowsky et al. 2008) method as implemented in the
astrodendro\footnote{\url{http://www.dendrograms.org/}} python
package. We set the minimum threshold intensity required to identify a
parent tree structure ({\it trunk}) to be $4\sigma$, where $\sigma$ is
the rms noise level in the continuum image prior to primary beam
correction, with typical value $\sigma \sim 0.2\:{\rm mJy
  \ beam}^{-1}$, except for C9 where $\sigma = 0.6\:{\rm mJy
  \ beam}^{-1}$ due to its large dynamic range.
  
For identification of nested substructures ({\it branches} and {\it
  leaves}), we require an additional $1\sigma$ increase in
intensity. Finally, we set a minimum area of half the synthesized beam
size for a leaf structure to be identified. These ``leaves'' are the
identified ``cores''. The parameters associated with these three
choices are the same as the fiducial choices of Paper I. We note that
Paper I carried out an extensive exploration of the effects of these
parameter choices on the derived CMF, which we do not carry out here,
rather focusing on the comparison of fiducial-method CMFs between the
IRDC clump and G286 protocluster environments.

While the dendrogram algorithm is our preferred fiducial method of
core identification, following Paper I, we will also
consider the effects of using the clumpfind algorithm (Williams et
al. 1994). The main differences of clumpfind are that it is
non-hierarchical, so that all the detected signal is apportioned
between the ``cores'', leading, in general, to more massive cores and
thus a more top-heavy CMF (see Paper I).

We note that one difference between our methodology compared to that
of Paper I is that our core identification is
done in images before primary beam correction. This is because our
observational data set consists of multiple individual pointings,
whereas that of Paper I is a mosaic of a single region, i.e.,
with a more uniform noise level. The result of this difference is that
our threshold levels that define cores vary depending on position in
the image. Our method of implementing completeness corrections,
described below, attempts to correct for this effect. Note, we
restrict core identification to the area within the FWHM primary beam in
each image.

\subsection{Core Mass Estimation}

We estimate core masses by assuming optically thin thermal emission
from dust, following the same assumptions adopted in Paper I.
The total mass surface density corresponding to a given
specific intensity of mm continuum emission is
\begin{eqnarray}
\Sigma_{\rm mm} & = & 0.369 \frac{F_\nu}{\rm mJy}\frac{(1\arcsec)^2}{\Omega} \frac{\lambda_{1.3}^3}{\kappa_{\nu,0.00638}}
 \nonumber\\
 & & \times  \left[{\rm exp}\left(0.553 T_{d,20}^{-1}
  \lambda_{1.3}^{-1}\right)-1\right]\:{\rm g\:cm^{-2}}\\
 &\rightarrow & 0.272 \frac{F_\nu}{\rm mJy}\frac{(1\arcsec)^2}{\Omega}\:{\rm g\:cm^{-2}}\nonumber,
\label{eq:Sigmamm}
\end{eqnarray}
where $F_{\nu}$ is the total integrated flux over solid angle
$\Omega$, $\kappa_{\nu,0.00638}\equiv\kappa_\nu/({\rm
  6.38\times10^{-3}\:cm^2\:g}^{-1})$ is the dust absorption
coefficient, $\lambda_{1.3}=\lambda/1.30\:{\rm mm}$ and
$T_{d,20}=T_d/20\:{\rm K}$ with $T_d$ being the dust temperature. To
obtain the above fiducial normalization of $\kappa_\nu$, we assume an
opacity per unit dust mass $\kappa_{\rm 1.3mm,d}=0.899\: {\rm cm^2
  g}^{-1}$ (moderately coagulated thin ice mantle model of Ossenkopf
\& Henning 1994), which then gives $\kappa_{\rm 1.3mm}= {\rm
  6.38\times10^{-3}\:cm^2\:g}^{-1}$ using a
gas-to-refractory-component-dust ratio of 141 (Draine 2011). The
numerical factor following the $\rightarrow$ in the final line shows
the fiducial case where $\lambda_{1.3}=1$ and $T_{d,20}=1$.

We note that even though temperatures in IRDCs are often measured to
be cooler than $20\:K$, e.g., $\sim15\:$K from studies using inversion
transitions of $\rm NH_3$ (e.g., Pillai et al. 2006; Sokolov et
al. 2017) or from multiwavelength sub-mm continuum emission maps
(e.g., Lim et al. 2016), we expect that most of the cores identified
in our images are protostellar cores that are internally heated to
somewhat higher temperatures. If temperatures of 15~K or 30~K were to
be adopted, then the mass estimates would differ by factors of 1.48
and 0.604, respectively.

\subsection{Core Flux Recovery and Completeness Corrections}\label{S:methods_cc}

Following Paper I, we estimate two correction factors
needed to estimate a ``true'' CMF from a ``raw'' observed CMF. The
first factor is the flux recovery fraction, $f_{\rm flux}$; the second
factor is the number recovery fraction, $f_{\rm num}$.

To evaluate these factors, artificial cores of a given mass (i.e.,
after primary beam correction) 
are inserted into each of the IRDC images, with three sources being
inserted at a given time at random locations within the primary beam
and this exercise repeated 50 times. This enables 150 experiments for each core mass. We note that the choice of random
placement within the primary beam is different from that adopted in
Paper I, which used the ACA-only image of the mosaic region as a
weighting factor for core placement.  We also note that our method
means that cores of a given mass that are placed near the edge of the
primary beam have smaller fluxes in the image and thus are harder to
detect. We explore a range of masses from $10^{-1}$ to
$10^{1.2}\:M_{\odot}$ with even spacing of 0.2 in log~$M$. We assume
the flux of the artificial cores has a gaussian distribution with the
shape of the synthesized beam. This is an approximation that is most
accurate in the limit of small, unresolved cores, which is where the
correction factors become most important.  The dendrogram algorithm is
run to determine if the cores are recovered and then the recovered
flux is compared to the true flux.

\begin{figure}[htbp]
\centering{\includegraphics[width=0.48\textwidth]{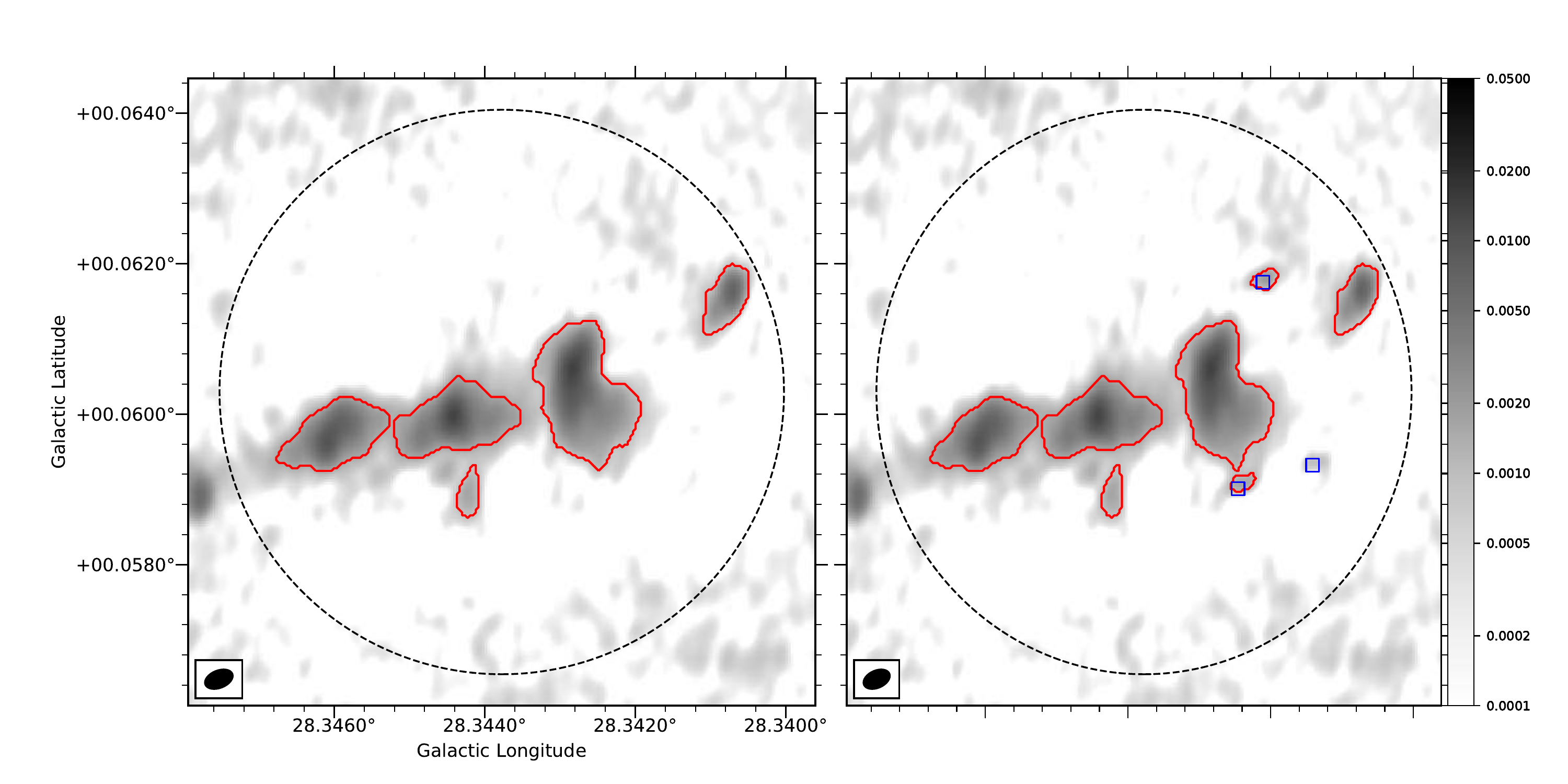}}
\caption{
An example of artificial core insertion and recovery to evaluate
completeness corrections for the C2 clump. {\it (a) Left:} Original
1.3~mm continuum image of the region (intensity scale in Jy beam$^{-1}$;
dashed circle shows FWHM of primary beam; synthesized beam shown in bottom
left), with boundaries of the identified cores shown in red. {\it (b) Right:} Same
as (a) but now after inserting three artificial cores of
$1.6\:M_{\odot}$ at random locations, with their centers marked by
blue squares. Cores identified by the dendrogram algorithm are again
marked with red contours: two out of three of the artificial cores are
found.
}\label{fig:eg}
\end{figure}

An example of this procedure is shown in Figure~\ref{fig:eg}.  We can
tell from the figure that whether a core of $\sim 1.6\:M_{\odot}$,
which has a peak flux of $\sim 10 \ \sigma$ at 5.0 kpc, can be
detected depends on its location within the filed of view, i.e., being
harder to detect near the edge of the primary beam, and also on the
local background. The local background can have two main
effects. First, if a faint core happens to be placed on an already
identified stronger core, then the artificial core is likely to be
undetected due to confusion. Second, if a faint core is placed on a
region of emission in the original image that was too faint to be
detected as a core, this increases the chances that the core will now
be recovered by the core finding algorithm. In this case its recovered
flux will have been artificially boosted by the presence of this
background emission, though the total recovered flux may still be less
than that inserted, e.g., due to the threshold criteria of core
finding algorithm.

The median value of the ratio of recovered to true flux defines
$f_{\rm flux}$, with this quantity being measured both as a function
of true flux (mass) and of recovered flux (mass). The ratio of the
actual number of cores recovered to the number inserted defines
$f_{\rm num}$.  The derived values of $f_{\rm flux}$ and $f_{\rm num}$
are presented below in \S\ref{S:results}.

\section{Results}\label{S:results}

\subsection{Continuum Images}\label{S:contim}

\begin{figure*}[htbp]
\centering{\includegraphics[width=18cm]{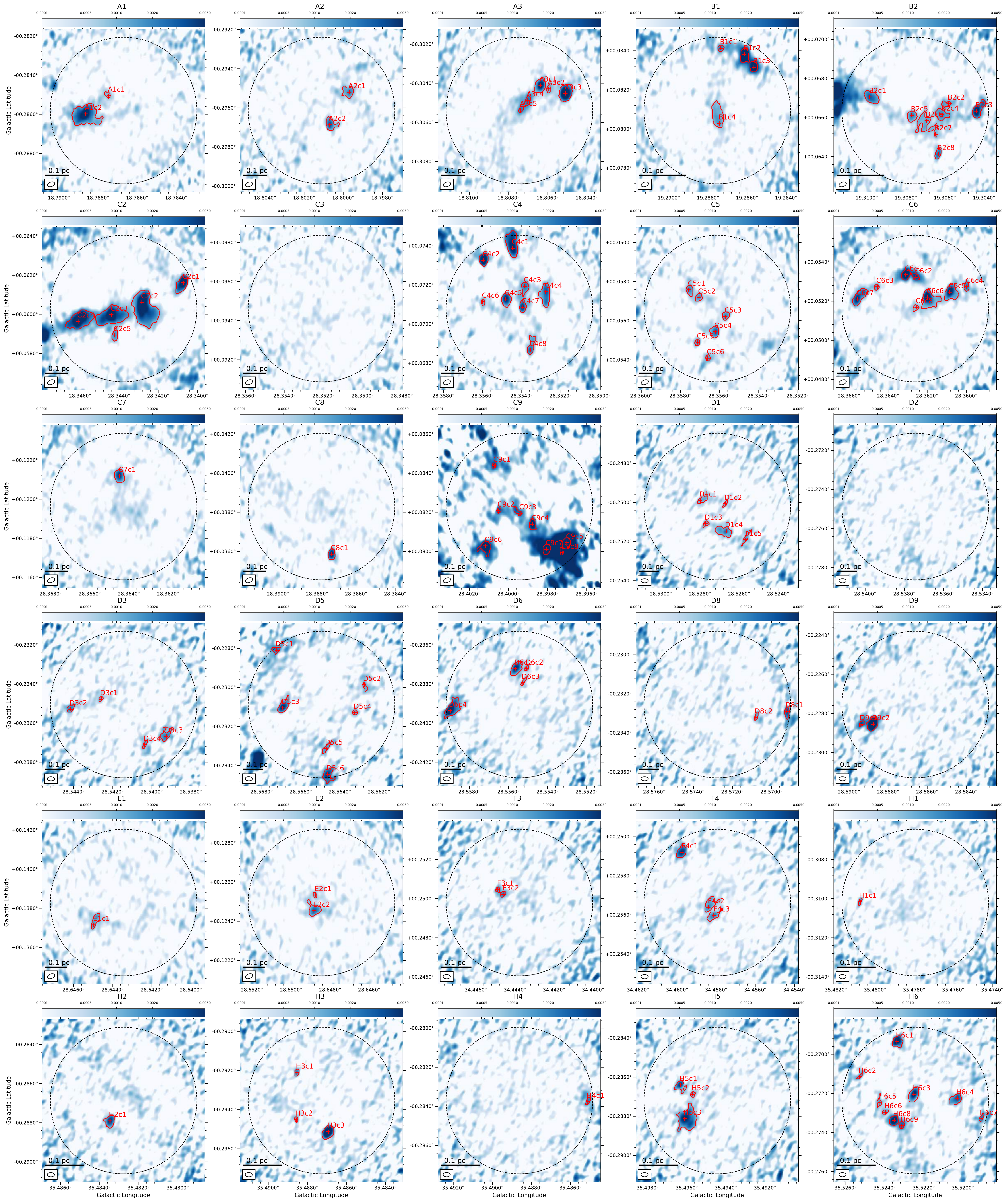}}
\caption{
1.3~mm continuum images of 30 pointings toward IRDC dense clumps
(colorbar in Jy beam$^{-1}$). The dotted circle in each panel denotes
the primary beam. The synthesized beam is shown in the bottom left
corner of each panel. The cores identified by the fiducial dendrogram
algorithm are marked on the images, with red contours showing ``leaf''
structures. 
}\label{fig:core} 
\end{figure*}

The continuum images of the 30 positions in the IRDCs, covering 32
clumps, are shown in Figure~\ref{fig:core}, together with the
identified cores (i.e., leaves from the dendrogram algorithm). The
size of the FWHM of the primary beam is shown with a dashed circle in
each image.

Overall we have identified 107 cores in these images. Note that we
only identify cores that are within the primary beam. Although there
may be true cores that show strong emission outside the primary beam,
as in B2 and C2, in most cases the noise outside is relatively high
and thus it is harder to identify cores of a given mass. We also note
that we identify cores in all the regions apart from C3 and D2. Cores
are named as, e.g., A1c1, A1c2, etc, in the region A1, with the
numbering order from higher to lower Galactic latitude.

The properties of the identified cores (after primary beam correction)
are listed in Table~\ref{tab:cores}. The masses range from $0.261
M_{\odot}$ to $178\:M_{\odot}$
(0.150 $M_\odot$ to 178
 $M_\odot$ without flux correction),
given our fiducial methods of mass estimation. The median radius of
the cores is $R_c\sim 0.02$~pc, with the radii evaluated as $R_c =
\sqrt{A/\pi}$, where $A$ is the projected area of the core. We
  then evaluate the mean mass surface density of the cores,
  $\Sigma_c\equiv M/A$, which have values $\gtrsim 0.3\:{\rm
    g\:cm}^{-2}$. This is consistent with expectations of the
  Turbulent Core Model of McKee \& Tan (2003) given that the mass
  surface densities of the IRDC clump environments are at about this
  level of $\sim0.3\:{\rm g\:cm}^{-2}$ (Butler \& Tan 2012). We also
  evaluate the mean H nuclei number density in the cores, $n_{{\rm H},c} 
  \equiv M_c/(\mu_{\rm H} V)$, where $\mu_{\rm H}=1.4 m_{\rm H}$
  is the mean mass per H assuming $n_{\rm He}=0.1 n_{\rm H}$ and
  $V=4\pi R_c^3/3$. The mean value of ${\rm log}_{10} (n_{{\rm H},c} /
  {\rm cm}^{-3})$ is 6.58, with a dispersion of 0.34.

From an inspection of the molecular line data of these regions, as
presented by Kong et al. (2017), we note that more than half of the
cores are associated with molecular line emission, e.g., $\rm
N_2D^+$(3-2), DCN(3-2), DCO$^{+}$(3-2), C$^{18}$O(2-1) and,
occasionally, SiO(5-4). However, only the latter of these transitions
is known to be a good tracer of outflows, especially from more massive
protostars. Analysis of the SiO emission will be presented in a
companion paper (Liu et al., in prep.).

\begin{deluxetable*}{ccccccccccc}
\tabletypesize{\scriptsize}
\tablecaption{Estimated physical parameters for 1.3~mm continuum cores \label{tab:cores} }
\tablewidth{18pt}
\tablehead{
  \colhead{Source} &  \colhead{$l$} &  \colhead{$b$} &  \colhead{$d$} &  \colhead{$I_{\rm peak}$} &  \colhead{$S_{\nu}$} &  \colhead{$M_{\rm c,raw}$}  &  \colhead{$M_{c}$}  &    \colhead{$R_{c}$} &  \colhead{$\Sigma_{c}$} &  \colhead{$n_{{\rm H},c}$} \\
 \colhead{}  &  \colhead{($\degr$)} &  \colhead{($\degr$)} &  \colhead{(kpc)} &  \colhead{(mJy $\rm beam^{-1}$)} &  \colhead{(mJy)} &  \colhead{$(M_{\odot})$} &  \colhead{$(M_{\odot})$}  & \colhead{(0.01 pc)} &   \colhead{(g $\rm cm^{-2}$)} &  \colhead{$(10^6 \ cm^{-3})$}  \\
 \vspace{-0.4cm}
 }
\startdata
A1c1 & 18.78746 & -0.28505 & 4.8 & 1.11 & 0.714 & 0.501 & 0.949 & 1.29 & 0.380 & 3.05 \\
A1c2 & 18.78864 & -0.28598 & 4.8 & 9.92 & 32.8 & 23.0 & 23.0 & 5.25 & 0.559 & 1.11 \\
A2c1 & 18.79969 & -0.29520 & 4.8 & 2.14 & 3.76 & 2.64 & 3.49 & 2.47 & 0.382 & 1.60 \\
A2c2 & 18.80070 & -0.29687 & 4.8 & 3.46 & 4.97 & 3.49 & 4.36 & 2.57 & 0.442 & 1.79 \\
A3c1 & 18.80637 & -0.30411 & 4.8 & 7.17 & 9.21 & 6.47 & 7.40 & 2.81 & 0.625 & 2.30 \\
A3c2 & 18.80596 & -0.30428 & 4.8 & 1.24 & 0.843 & 0.592 & 1.12 & 1.32 & 0.428 & 3.36 \\
A3c3 & 18.80509 & -0.30452 & 4.8 & 14.24 & 23.4 & 16.5 & 16.5 & 3.33 & 0.992 & 3.09 \\
A3c4 & 18.80703 & -0.30487 & 4.8 & 2.51 & 2.64 & 1.86 & 2.69 & 1.68 & 0.635 & 3.92 \\
A3c5 & 18.80738 & -0.30536 & 4.8 & 2.31 & 1.38 & 0.971 & 1.73 & 1.25 & 0.741 & 6.15 \\
B1c1 & 19.28735 & 0.08413 & 2.4 & 2.24 & 1.57 & 0.277 & 0.474 & 0.710 & 0.637 & 9.36 \\
B1c2 & 19.28614 & 0.08382 & 2.4 & 17.10 & 23.8 & 4.18 & 4.47 & 1.27 & 1.85 & 15.1 \\
B1c3 & 19.28565 & 0.08316 & 2.4 & 11.08 & 12.5 & 2.20 & 2.46 & 1.04 & 1.51 & 15.0 \\
B1c4 & 19.28742 & 0.08028 & 2.4 & 1.69 & 7.20 & 1.26 & 1.52 & 1.87 & 0.291 & 1.62 \\
B2c1 & 19.30985 & 0.06706 & 2.4 & 3.37 & 7.22 & 1.27 & 1.52 & 1.47 & 0.472 & 3.34 \\
B2c2 & 19.30582 & 0.06671 & 2.4 & 1.54 & 1.67 & 0.293 & 0.493 & 0.840 & 0.466 & 5.75 \\
B2c3 & 19.30440 & 0.06633 & 2.4 & 8.84 & 12.3 & 2.16 & 2.42 & 1.43 & 0.791 & 5.75 \\
B2c4 & 19.30614 & 0.06615 & 2.4 & 1.98 & 4.44 & 0.780 & 1.00 & 1.32 & 0.387 & 3.05 \\
B2c5 & 19.30770 & 0.06612 & 2.4 & 2.20 & 3.19 & 0.561 & 0.781 & 1.15 & 0.398 & 3.60 \\
B2c6 & 19.30694 & 0.06584 & 2.4 & 1.35 & 4.66 & 0.818 & 1.04 & 1.56 & 0.287 & 1.91 \\
B2c7 & 19.30648 & 0.06515 & 2.4 & 1.15 & 1.76 & 0.309 & 0.512 & 0.990 & 0.349 & 3.65 \\
B2c8 & 19.30634 & 0.06414 & 2.4 & 2.69 & 2.50 & 0.440 & 0.660 & 0.890 & 0.560 & 6.54 \\
C2c1 & 28.34072 & 0.06161 & 5.0 & 12.07 & 16.9 & 12.9 & 14.0 & 3.12 & 0.962 & 3.19 \\
C2c2 & 28.34284 & 0.06061 & 5.0 & 14.05 & 63.6 & 48.5 & 48.5 & 6.57 & 0.750 & 1.18 \\
C2c3 & 28.34440 & 0.05998 & 5.0 & 13.19 & 41.8 & 31.9 & 31.9 & 5.31 & 0.755 & 1.47 \\
C2c4 & 28.34610 & 0.05963 & 5.0 & 12.74 & 43.4 & 33.1 & 33.1 & 4.80 & 0.960 & 2.07 \\
C2c5 & 28.34423 & 0.05894 & 5.0 & 1.77 & 2.02 & 1.54 & 2.39 & 1.98 & 0.408 & 2.14 \\
C4c1 & 28.35446 & 0.07388 & 5.0 & 6.73 & 22.1 & 16.8 & 16.8 & 4.01 & 0.700 & 1.81 \\
C4c2 & 28.35596 & 0.07326 & 5.0 & 12.77 & 12.7 & 9.65 & 10.7 & 2.50 & 1.15 & 4.76 \\
C4c3 & 28.35384 & 0.07194 & 5.0 & 2.31 & 3.07 & 2.34 & 3.23 & 2.13 & 0.477 & 2.32 \\
C4c4 & 28.35276 & 0.07166 & 5.0 & 3.31 & 9.02 & 6.87 & 7.93 & 3.49 & 0.436 & 1.30 \\
C4c5 & 28.35481 & 0.07128 & 5.0 & 5.68 & 7.23 & 5.51 & 6.55 & 2.67 & 0.614 & 2.39 \\
C4c6 & 28.35599 & 0.07114 & 5.0 & 1.31 & 0.667 & 0.509 & 0.941 & 1.21 & 0.431 & 3.70 \\
C4c7 & 28.35394 & 0.07086 & 5.0 & 4.58 & 4.52 & 3.45 & 4.39 & 2.16 & 0.627 & 3.01 \\
C4c8 & 28.35356 & 0.06867 & 5.0 & 2.96 & 4.07 & 3.10 & 4.02 & 2.55 & 0.413 & 1.68 \\
C5c1 & 28.35757 & 0.05759 & 5.0 & 2.02 & 2.20 & 1.68 & 2.55 & 1.99 & 0.428 & 2.23 \\
C5c2 & 28.35705 & 0.05718 & 5.0 & 1.67 & 1.48 & 1.13 & 1.91 & 1.73 & 0.424 & 2.53 \\
C5c3 & 28.35570 & 0.05621 & 5.0 & 1.99 & 1.75 & 1.33 & 2.15 & 1.83 & 0.431 & 2.45 \\
C5c4 & 28.35622 & 0.05544 & 5.0 & 2.87 & 3.77 & 2.88 & 3.79 & 2.41 & 0.438 & 1.89 \\
C5c5 & 28.35712 & 0.05489 & 5.0 & 1.87 & 1.04 & 0.794 & 1.47 & 1.36 & 0.533 & 4.08 \\
C5c6 & 28.35660 & 0.05409 & 5.0 & 1.52 & 0.871 & 0.664 & 1.23 & 1.28 & 0.502 & 4.07 \\
C6c1 & 28.36310 & 0.05336 & 5.0 & 11.38 & 12.2 & 9.30 & 10.4 & 2.27 & 1.35 & 6.18 \\
C6c2 & 28.36258 & 0.05322 & 5.0 & 4.38 & 3.95 & 3.01 & 3.93 & 1.66 & 0.956 & 5.98 \\
C6c3 & 28.36456 & 0.05273 & 5.0 & 1.64 & 0.852 & 0.649 & 1.20 & 1.27 & 0.500 & 4.09 \\
C6c4 & 28.35998 & 0.05273 & 5.0 & 2.38 & 1.76 & 1.34 & 2.15 & 1.58 & 0.579 & 3.81 \\
C6c5 & 28.36085 & 0.05246 & 5.0 & 7.80 & 11.5 & 8.77 & 9.86 & 3.32 & 0.597 & 1.86 \\
C6c6 & 28.36199 & 0.05221 & 5.0 & 9.28 & 15.5 & 11.8 & 13.0 & 3.96 & 0.553 & 1.45 \\
C6c7 & 28.36557 & 0.05211 & 5.0 & 5.19 & 9.54 & 7.28 & 8.34 & 3.12 & 0.573 & 1.91 \\
C6c8 & 28.36255 & 0.05169 & 5.0 & 1.11 & 0.774 & 0.590 & 1.09 & 1.44 & 0.352 & 2.53 \\
C7c1 & 28.36448 & 0.12119 & 5.0 & 4.91 & 6.34 & 4.83 & 5.85 & 2.69 & 0.539 & 2.08 \\
C8c1 & 28.38725 & 0.03586 & 5.0 & 5.85 & 5.09 & 3.88 & 4.84 & 2.11 & 0.724 & 3.55 \\
C9c1 & 28.40073 & 0.08438 & 5.0 & 4.25 & 1.88 & 1.43 & 2.26 & 1.11 & 1.23 & 11.5 \\
C9c2 & 28.40052 & 0.08209 & 5.0 & 4.09 & 1.77 & 1.35 & 2.17 & 1.16 & 1.08 & 9.64 \\
C9c3 & 28.39941 & 0.08195 & 5.0 & 3.65 & 3.23 & 2.46 & 3.36 & 1.63 & 0.845 & 5.37 \\
C9c4 & 28.39878 & 0.08139 & 5.0 & 8.38 & 8.36 & 6.37 & 7.42 & 2.07 & 1.16 & 5.78 \\
C9c5 & 28.39701 & 0.08045 & 5.0 & 196.87 & 233 & 178 & 178 & 2.22 & 24.0 & 112 \\
C9c6 & 28.40118 & 0.08028 & 5.0 & 11.94 & 18.2 & 13.9 & 15.0 & 2.74 & 1.34 & 5.05 \\
C9c7 & 28.39806 & 0.08011 & 5.0 & 28.96 & 33.5 & 25.6 & 25.6 & 2.08 & 3.95 & 19.7 \\
C9c8 & 28.39726 & 0.07993 & 5.0 & 85.31 & 51.3 & 39.1 & 39.1 & 1.28 & 16.0 & 130 \\
D1c1 & 28.52798 & -0.24990 & 5.7 & 1.30 & 1.92 & 1.90 & 2.89 & 2.24 & 0.385 & 1.78 \\
D1c2 & 28.52670 & -0.25007 & 5.7 & 1.25 & 0.603 & 0.598 & 1.02 & 1.28 & 0.416 & 3.38 \\
D1c3 & 28.52771 & -0.25108 & 5.7 & 1.29 & 0.890 & 0.882 & 1.50 & 1.52 & 0.433 & 2.95 \\
D1c4 & 28.52666 & -0.25146 & 5.7 & 2.47 & 5.39 & 5.34 & 6.37 & 3.34 & 0.383 & 1.19 \\
D1c5 & 28.52569 & -0.25191 & 5.7 & 1.66 & 1.77 & 1.75 & 2.74 & 2.02 & 0.451 & 2.32 \\
D3c1 & 28.54259 & -0.23477 & 5.7 & 1.27 & 0.597 & 0.591 & 1.01 & 1.26 & 0.422 & 3.46 \\
D3c2 & 28.54416 & -0.23529 & 5.7 & 2.17 & 2.17 & 2.15 & 3.15 & 1.93 & 0.565 & 3.04 \\
D3c3 & 28.53926 & -0.23668 & 5.7 & 2.37 & 4.2 & 4.16 & 5.18 & 2.74 & 0.463 & 1.75 \\
D3c4 & 28.54037 & -0.23710 & 5.7 & 1.59 & 1.02 & 1.01 & 1.71 & 1.47 & 0.528 & 3.72 \\
D5c1 & 28.56724 & -0.22810 & 5.7 & 2.43 & 2.64 & 2.61 & 3.61 & 1.93 & 0.649 & 3.49 \\
D5c2 & 28.56276 & -0.22987 & 5.7 & 1.35 & 0.988 & 0.979 & 1.66 & 1.53 & 0.471 & 3.18 \\
D5c3 & 28.56693 & -0.23105 & 5.7 & 5.69 & 7.96 & 7.89 & 8.96 & 3.13 & 0.612 & 2.03 \\
D5c4 & 28.56324 & -0.23129 & 5.7 & 1.32 & 0.799 & 0.792 & 1.35 & 1.42 & 0.448 & 3.27 \\
D5c5 & 28.56470 & -0.23313 & 5.7 & 1.69 & 1.77 & 1.76 & 2.74 & 1.95 & 0.483 & 2.57 \\
D5c6 & 28.56463 & -0.23445 & 5.7 & 4.89 & 8.90 & 8.82 & 9.91 & 3.09 & 0.695 & 2.33 \\
D6c1 & 28.55565 & -0.23721 & 5.7 & 5.47 & 8.73 & 8.65 & 9.74 & 3.22 & 0.628 & 2.02 \\
D6c2 & 28.55507 & -0.23721 & 5.7 & 1.46 & 0.658 & 0.652 & 1.11 & 1.23 & 0.488 & 4.11 \\
D6c3 & 28.55527 & -0.23794 & 5.7 & 1.18 & 0.645 & 0.639 & 1.09 & 1.34 & 0.407 & 3.16 \\
D6c4 & 28.55899 & -0.23936 & 5.7 & 10.89 & 19.8 & 19.6 & 19.6 & 3.99 & 0.823 & 2.14 \\
D8c1 & 28.56923 & -0.23289 & 5.7 & 3.59 & 3.70 & 3.67 & 4.68 & 2.03 & 0.763 & 3.91 \\
D8c2 & 28.57080 & -0.23321 & 5.7 & 1.41 & 0.851 & 0.843 & 1.43 & 1.39 & 0.495 & 3.69 \\
D9c1 & 28.58939 & -0.22855 & 5.7 & 3.94 & 2.40 & 2.38 & 3.38 & 1.46 & 1.06 & 7.55 \\
D9c2 & 28.58877 & -0.22855 & 5.7 & 22.55 & 28.5 & 28.3 & 28.3 & 3.13 & 1.93 & 6.39 \\
E1c1 & 28.64497 & 0.13715 & 5.1 & 1.63 & 2.69 & 2.14 & 2.98 & 2.37 & 0.356 & 1.56 \\
E2c1 & 28.64876 & 0.12534 & 5.1 & 1.22 & 0.511 & 0.405 & 0.704 & 1.16 & 0.352 & 3.15 \\
E2c2 & 28.64883 & 0.12454 & 5.1 & 2.85 & 4.69 & 3.72 & 4.59 & 2.89 & 0.368 & 1.32 \\
F3c1 & 34.44489 & 0.25046 & 3.7 & 1.95 & 0.979 & 0.409 & 0.661 & 0.870 & 0.588 & 7.03 \\
F3c2 & 34.44461 & 0.25022 & 3.7 & 2.36 & 1.44 & 0.602 & 0.973 & 1.01 & 0.635 & 6.51 \\
F4c1 & 34.45975 & 0.25920 & 3.7 & 4.91 & 7.55 & 3.15 & 3.60 & 1.85 & 0.706 & 3.97 \\
F4c2 & 34.45840 & 0.25639 & 3.7 & 1.88 & 4.08 & 1.71 & 2.16 & 2.05 & 0.344 & 1.74 \\
F4c3 & 34.45812 & 0.25597 & 3.7 & 2.23 & 3.19 & 1.33 & 1.78 & 1.74 & 0.391 & 2.32 \\
H1c1 & 35.48076 & -0.31016 & 2.9 & 1.74 & 0.783 & 0.201 & 0.348 & 0.630 & 0.592 & 9.80 \\
H2c1 & 35.48347 & -0.28791 & 2.9 & 4.90 & 6.58 & 1.69 & 1.97 & 1.49 & 0.595 & 4.15 \\
H3c1 & 35.48853 & -0.29211 & 2.9 & 2.21 & 1.29 & 0.330 & 0.540 & 0.800 & 0.565 & 7.33 \\
H3c2 & 35.48856 & -0.29451 & 2.9 & 1.39 & 0.586 & 0.150 & 0.261 & 0.620 & 0.455 & 7.62 \\
H3c3 & 35.48693 & -0.29513 & 2.9 & 20.70 & 22.8 & 5.86 & 6.17 & 1.68 & 1.46 & 8.98 \\
H4c1 & 35.48512 & -0.28377 & 2.9 & 2.33 & 1.46 & 0.374 & 0.603 & 0.760 & 0.707 & 9.71 \\
H5c1 & 35.49632 & -0.28640 & 2.9 & 4.93 & 5.34 & 1.37 & 1.65 & 1.42 & 0.543 & 3.96 \\
H5c2 & 35.49570 & -0.28688 & 2.9 & 1.36 & 0.732 & 0.188 & 0.326 & 0.700 & 0.443 & 6.55 \\
H5c3 & 35.49611 & -0.28813 & 2.9 & 6.12 & 24.9 & 6.39 & 6.70 & 2.74 & 0.599 & 2.27 \\
H6c1 & 35.52338 & -0.26935 & 2.9 & 8.98 & 10.7 & 2.74 & 3.03 & 1.46 & 0.955 & 6.80 \\
H6c2 & 35.52529 & -0.27115 & 2.9 & 1.79 & 0.867 & 0.222 & 0.386 & 0.640 & 0.625 & 10.1 \\
H6c3 & 35.52251 & -0.27205 & 2.9 & 7.24 & 9.03 & 2.31 & 2.60 & 1.64 & 0.645 & 4.08 \\
H6c4 & 35.52029 & -0.27226 & 2.9 & 3.54 & 6.03 & 1.55 & 1.82 & 1.52 & 0.530 & 3.63 \\
H6c5 & 35.52425 & -0.27247 & 2.9 & 1.33 & 1.26 & 0.322 & 0.529 & 0.910 & 0.423 & 4.81 \\
H6c6 & 35.52397 & -0.27296 & 2.9 & 1.51 & 0.921 & 0.236 & 0.407 & 0.760 & 0.478 & 6.56 \\
H6c7 & 35.51908 & -0.27330 & 2.9 & 2.33 & 1.41 & 0.363 & 0.587 & 0.740 & 0.726 & 10.2 \\
H6c8 & 35.52352 & -0.27337 & 2.9 & 7.96 & 9.86 & 2.53 & 2.82 & 1.31 & 1.10 & 8.74 \\
H6c9 & 35.52314 & -0.27365 & 2.9 & 3.05 & 2.46 & 0.631 & 0.892 & 0.850 & 0.820 & 9.98 \\
\enddata
\tablecomments{
$M_c$ is the mass estimate after flux correction, which
equals the raw, uncorrected mass estimate ($M_{c \rm,raw}$) multiplied
by the value of $f_{\rm flux}^{-1}$ appropriate for $M_c$.  This corrected
mass is then used for the estimates of $\Sigma_c$ and $n_{{\rm H},c}$.}
\end{deluxetable*}

\subsection{Core Mass Function}
\begin{figure}[htbp]
\centering{\includegraphics[width=0.48\textwidth]{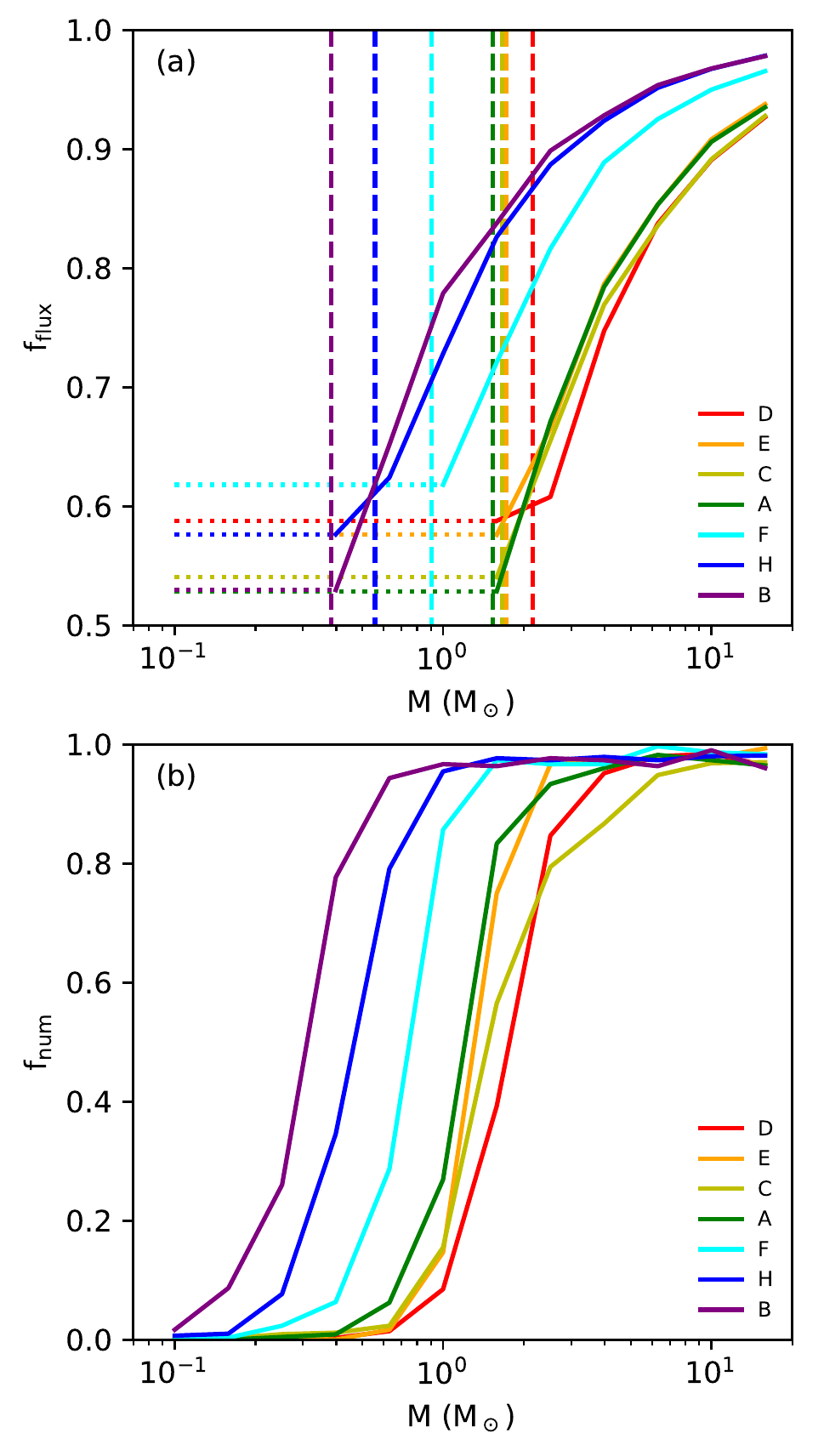}}
\caption{
{\it (a) Top:} Average (median) flux recovery fractions, $f_{\rm
  flux}$, versus core mass, $M$, for the dendrogram core finding
method as applied to each of the seven IRDCs (solid lines; see
legend). Note that our method assumes a constant value of $f_{\rm
  flux}$ (dotted portion of lines)  once an effective minimum is reached as $M$ is reduced (see
text). Vertical dashed lines show the mass corresponding to a core
that has a flux level of $4\sigma$ at the position of half the beam size, which represents one of
the detection threshold criteria, assuming its flux distribution is shaped as the beam. Note that the legend is ordered by
cloud distance: IRDC B is the nearest; IRDC D is the most distant.
{\it (b) Bottom:} Mean value of the number recovery fraction, $f_{\rm
  num}$, versus core mass, $M$, for the dendrogram core finding method
as applied to each of the seven IRDCs (solid lines; see legend).
}\label{fig:corr}
\end{figure}

\begin{figure}[htbp]
\centering{\includegraphics[width=0.48\textwidth]{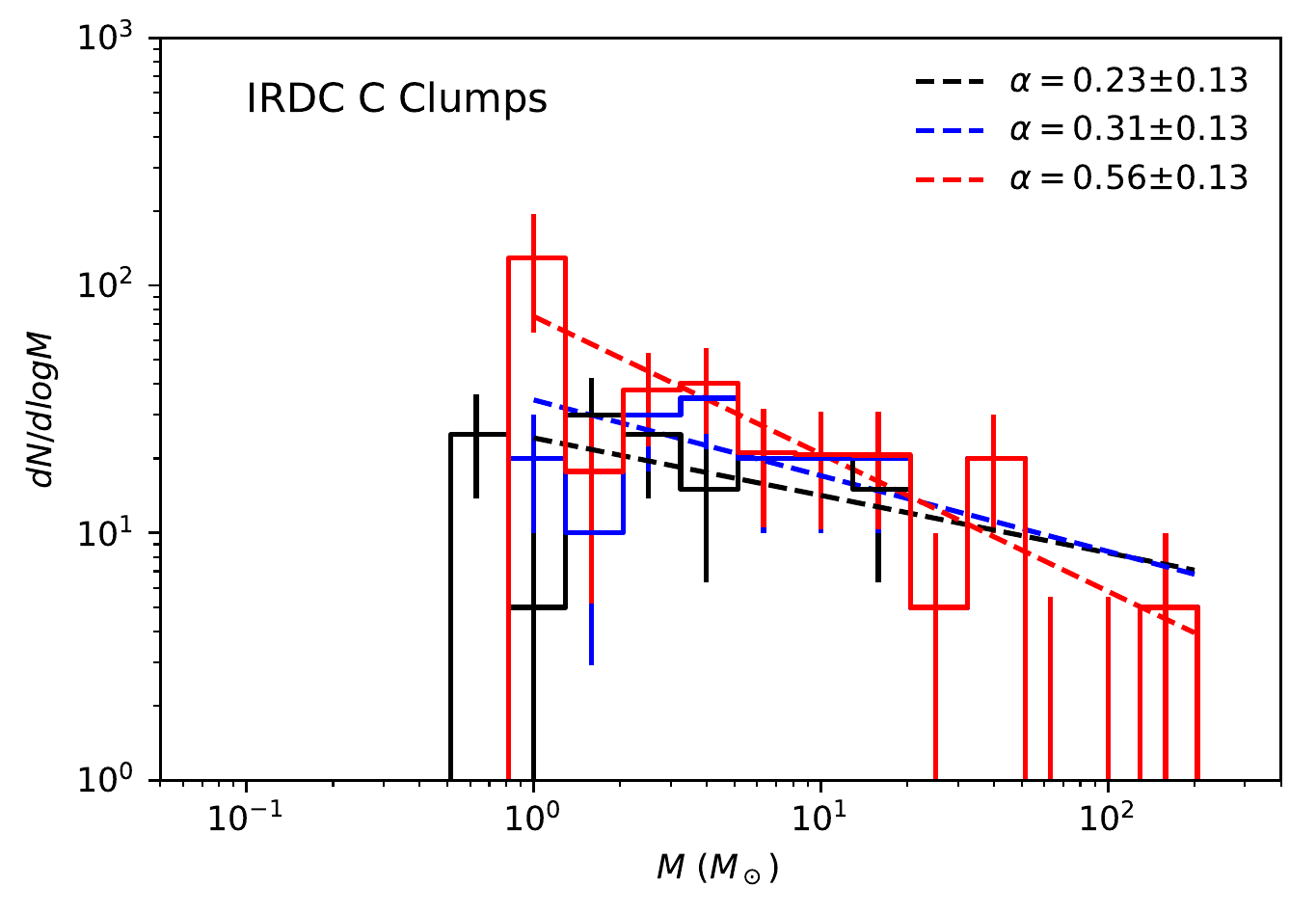}}
\caption{
The dendrogram-derived combined CMF of the seven clumps observed in
IRDC C. The black histogram shows the original, ``raw'' CMF. The blue
histogram shows the CMF after flux correction and the red histogram
shows the final, ``true'' CMF after then applying number recovery
fraction correction. The error bars show Poisson counting errors. The
black, blue and red dashed lines show the best power law fit results
for the high-mass end ($M\geq 0.79\:M_{\odot}$) of these CMFs,
respectively.}\label{fig:cmf_C}
\end{figure}

\begin{figure*}[htbp]
\centering{\includegraphics[width=18cm]{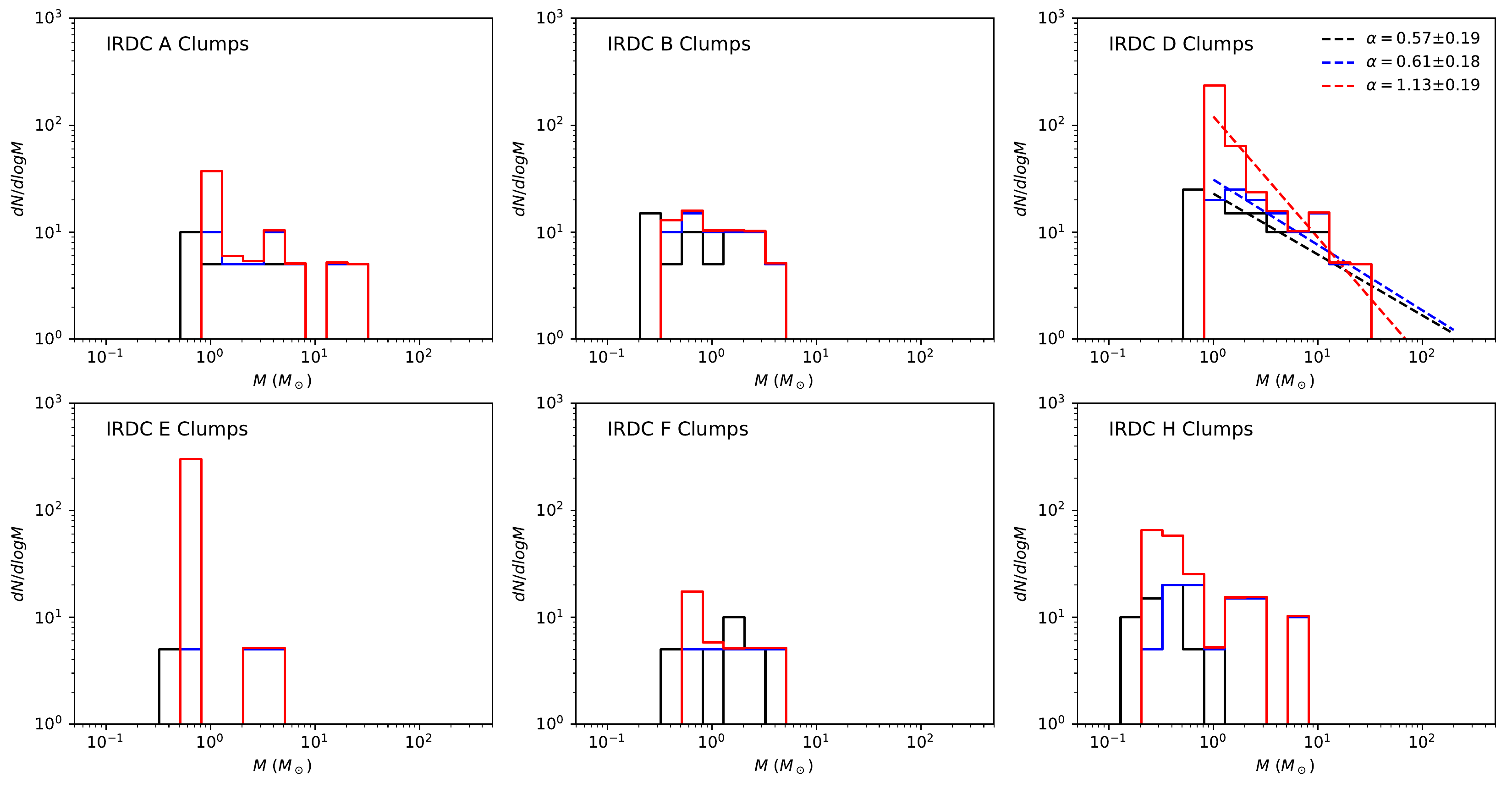}}
\caption{
Similar to Figure~\ref{fig:cmf_C} for IRDC C, the raw (black),
flux-corrected (blue) and true (red) CMFs are shown here for IRDCs A,
B, D, E, F and H. The black, blue and red dashed lines show the best
power law fit results for the high-mass end ($M\geq 0.79\:M_{\odot}$)
of the CMF in IRDC D. Other IRDCs are not fit, given their relatively
small number of cores.}\label{fig:cmf_indi}
\end{figure*}

As described in \S\ref{S:methods_cc}, we have estimated flux
correction, $f_{\rm flux}$, functions for all the observed regions and
these are shown in Figure~\ref{fig:corr}a for the seven IRDCs. Here
the values shown are the median of the results for each IRDC in each
mass bin (excluding values $f_{\rm flux}>1$, which we attribute to
false assignments; and extrapolating with constant values at the
low-mass end once an effective minimum is reached in the distribution:
at even smaller values of $M$, the median $f_{\rm flux}$ is seen to
rise, which we attribute to false assignment to weak image feature,
including noise fluctuations). Similar to the results of Paper I for
G286, our estimated values of $f_{\rm flux}$ rise from $\sim0.5$ to
0.6 for $M\lesssim 1\:M_\odot$ towards close to unity for $M\gtrsim$
several $M_\odot$. The curves are shifted to lower masses for the most
nearby IRDCs. Figure~\ref{fig:corr}a also shows for each IRDC the
  masses corresponding to a core that has a flux level of $4\sigma$ at
  the position of half the beam size, which represents one of the
  detection threshold criteria (in this case the most stringent),
  assuming its flux distribution is shaped as the beam. These mass
  detection limits range from about 0.4$\:M_\odot$ to 2$\:M_\odot$,
  depending on the distance to the cloud. However, we note that these
  are only approximate limits, since, e.g., the core shape may not be
  exactly the same as the beam. In particular, less centrally peaked
  cores will be able to satisfy the area threshold condition at a
  lower mass.  

As also described in \S\ref{S:methods_cc}, we then derive the number
recovery fraction, $f_{\rm num}$, for the observed regions, again
averaging for each IRDC (Fig.~\ref{fig:corr}b). These rise steeply
from near zero to near unity as $M$ increases from $\sim0.2$ to
$1\:M_\odot$, depending on the distance to the IRDC.

Recall that overall we have identified 107 cores in the seven
IRDCs. Cloud C contains the most (37), followed by cloud D (23) and
cloud H (18). We will first derive the CMFs for each IRDC separately
and then combine them.

The raw (uncorrected) CMF of IRDC C is shown by the black histogram in
Figure~\ref{fig:cmf_C}. The mass binning has been chosen to match that
used in Paper I, i.e., five bins per dex, with a bin centered on
$1\:M_\odot$ (and thus also on $10\:M_\odot$ and $100\:M_\odot$,
etc). The error bars on each bin indicate $N^{1/2}$ Poisson counting
uncertainties. The CMF after flux correction is shown by the blue
histogram: note that cores in the lowest mass bin of the raw CMF are
all shifted to higher mass bins. Finally, the number correction is
applied to the flux-corrected CMF to derive the final, ``true'' CMF,
shown by the red histogram. Note, its error bars are assumed to be the
same fractional size as those found for the blue histogram, i.e., the
Poisson errors from this distribution, with no allowance for any
additional systematic uncertainty in $f_{\rm num}$. Thus these
uncertainties should be treated with caution, i.e., they likely
underestimate the true uncertainties.

Following Paper I, we first carry out ``simple'' power law fitting to
CMFs starting from the $1\:M_\odot$ bin, i.e., for
$M\gtrsim0.79\:M_\odot$. This fitting minimizes differences in the log
of $dN/d{\rm log}M$, normalized using the asymmetric Poisson
errors. For empty bins, to treat these as effective upper limits, we
assume the point is 1 dex lower than the level if the bin had 1 data
point and set the upper error bar such that it reaches up to the level
if there were 1 data point. For bins that have 1 data point, the lower
error bar extends down by 1 dex rather than to minus infinity. As with
Paper I, we have verified that the global results are insensitive to
the details of how empty bins are treated.

We also apply a maximum likelihood estimation (MLE) method to estimate
the power law index (Newman et al. 2005). Let $p(x) dx$ be the
fraction of cores with mass between $x$ and $x + dx$. Then
$p(x) = Cx^{-(\alpha+1)}$
and $\alpha$ is estimated as
\begin{equation}
\alpha = n \left[ \sum_{i=1}^{n} {\rm ln} \frac{x_{i}}{x_{\rm min}} \right]^{-1}
\end{equation}
with an uncertainty (confidence interval)
\begin{equation}
\sigma = \frac{\alpha}{\sqrt{n}}.
\end{equation}
Here $x_{\rm min}$ is the starting mass of the power law, $x_{i}$ is
the mass of each core with mass above $x_{\rm min}$ and $n$ is the
number of such cores. We note that this estimate is
valid assuming the upper limit (if any) of the distribution
is much larger than $x_{\rm min}$. Note also, our fiducial results involve CMFs that
have been corrected in logarithmic bins for flux and number
incompleteness, so these are used to generate synthetic populations of cores, to which the MLE analysis
method is then applied.
We generate the corresponding number of random masses uniformly distributed in each mass bin and apply the MLE method. We repeat this for 50 times and then derive the median $\alpha$ and confidence interval $\sigma$.

For IRDC C, with simple power law fitting we derive a value of
$\alpha=0.56 \pm 0.13$ for the true CMF. The raw and flux-corrected
CMFs had power law indices of 0.23 and 0.31, respectively, so we see
the effects of these corrections has been to steepen the upper end
slope of the CMF, as expected. For the MLE method we find 
$\alpha=0.48\pm0.08$, $0.49\pm0.08$ and
$0.75 \pm 0.09$ for the raw, flux-corrected and ``true" CMF.
The slopes derived from the MLE method are slightly steeper than those derived from the linear fitting method within 1.5 combined $\sigma$.

\begin{figure}[htbp]
\centering{\includegraphics[width=0.48\textwidth]{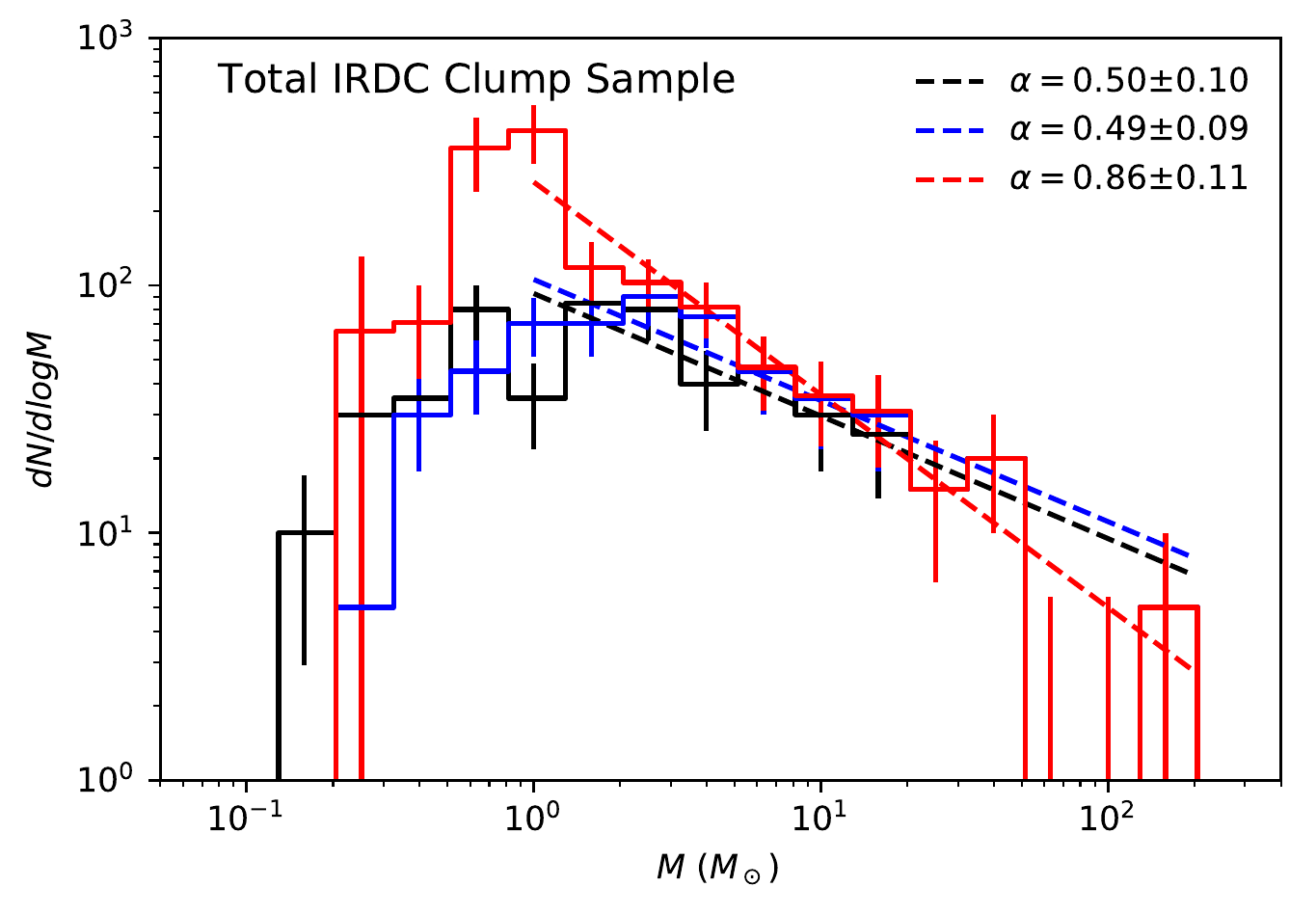}}
\caption{
Combined dendrogram-derived CMF from observations of 30 positions
covering dense clumps within seven IRDCs. The black histogram shows
the original, ``raw'' CMF. The blue histogram shows the CMF after flux
correction and the red histogram shows the final, ``true'' CMF after
then applying number recovery fraction correction. The error bars show
Poisson counting errors. The black, blue and red dashed lines show the
best power law fit results for the high-mass end ($M\geq
0.79\:M_{\odot}$) of these CMFs, respectively.}\label{fig:cmf_all}
\end{figure}

In Figure~\ref{fig:cmf_indi} we show the equivalent CMFs for the six
other IRDCs, most of which are very sparsely sampled. We also carry
out power-law fitting for IRDC D (23 cores). From simple power law
fitting we derive a value of $\alpha=1.13 \pm 0.19$ for the true
CMF. This is significantly steeper than the result for IRDC C,
however, it driven mostly by the lowest mass bin, i.e.,
$\sim1\:M_\odot$, for which the completeness correction is about a
factor of 10. Due to potential uncertainties associated with this
correction, as we discuss below, we will consider CMFs down to two
mass thresholds, i.e., cases of including and excluding this mass bin.

Next, in Figure~\ref{fig:cmf_all}, we show the combined CMFs for the
entire sample of seven IRDCs. The raw, flux-corrected and ``true''
CMFs (black, blue and red histograms, respectively) are obtained by
simple addition of the equivalent CMFs for each individual IRDC. Note
that the Poisson errors are now reduced. Note also, however, there are
still two empty bins near $100\:M_\odot$ and only one core more
massive that this. At the low-mass end, the CMFs of the seven
individual IRDCs all have detections down to or below the bin centered
on $M=1\:M_\odot$, which is approximately the detection threshold of
Cloud D, the farthest cloud.

For the raw, flux-corrected and ``true'' CMFs, with simple fitting we
then derive power law indices for $M>0.79\:M_\odot$ of
$\alpha=0.50\pm0.10$, $0.49\pm0.09$ and $0.86\pm0.11$, respectively. For MLE, we derive
$\alpha=0.61\pm0.07$, $0.63\pm0.07$ and $1.02\pm0.08$ for these three cases, respectively. Again, the slopes derived from the MLE method are slightly steeper than those derived from the linear fitting method within 1.5 combined $\sigma$.
If we only fit to the true CMF starting from the next bin above $1\:M_\odot$ (i.e., allowing
for the possibility that IRDC D is artificially distorting the low-end
CMF), then we derive $\alpha=0.70\pm0.13$ for the true CMF. The MLE analysis yields $\alpha=0.83\pm0.09$.

While we prefer the dendrogram algorithm as our fiducial method of
identifying cores, since it is a hierarchical method that we consider
better at separating cores from a surrounding background clump
environment (see \S\ref{S:methods_ci} and Paper I), for completeness
we also evaluate the CMF as derived from the clumpfind algorithm. With
the fiducial parameters (i.e., a $4\sigma$ noise threshold, $3\sigma$
step size, minimum area of 0.5 beams; see Paper I), we find 120 cores
with masses from 0.150 to $286\:M_{\odot}$. After flux and number
recovery corrections on each IRDC, for the combined ``true'' CMF we
derive a high-mass end ($M>0.79\:M_\odot$) power law index of $\alpha
= 0.86 \pm 0.11$ with simple fitting and $1.02\pm0.08$ with MLE
fitting. The first of these values is coincidentally the same (within
the first two significant figures) as that derived from the dendrogram
analysis. These results indicate that for our {\it ALMA} observations
of IRDC clumps, the resulting core properties are not that sensitive
to whether dendrogram or clumpfind is used as the identification
algorithm. This contrasts with the results of Paper I for G286 (for
the case of 1.5\arcsec resolution), which found a value of
$\alpha=1.12\pm0.18$ for dendrogram and $\alpha=0.49\pm0.12$ for
clumpfind. We expect that this difference is due, at least in part, to
the observation of G286 utilizing both the 12-m and 7-m arrays, so
that a larger range of scales are recovered. Thus more emission from
the surrounding protocluster clump material is detected in G286,
readily apparent from Figures 1 \& 2 of Paper I, in comparison to our
images of the IRDC clumps (Fig.~\ref{fig:core}).  Since most of the
larger-scale emission is resolved out in our IRDC observations (an
approximate comparison with BGPS data of the clumps assuming a dust spectra index
of 3.5 finds typical flux recovery of $\sim20\%$ [see \S\ref{S:obs}]), one
then expects clumpfind results to be closer to those derived from
dendrogram.

 We examine whether the CMF we measure in IRDC environments is
  consistent with a Sapleter distribution ($\alpha=1.35$).  We can
  already infer from our measurements of $\alpha=0.70\pm0.13$ (or with
  MLE $\alpha=0.83\pm0.09$) for the true CMF at $M>1.26\:M_\odot$, that
  the result differs from Salpeter by about $5.0\sigma$ (or
  $5.8\sigma$ for MLE). However, it is not known if the uncertainties
  in these parameters, especially given systematic uncertainties, will
  follow a simple gaussian distribution. More generally we compare the
  IRDC core population (including allowance for completeness
  corrections) with an idealized large (e.g., $1,000$, but result is
  independent of this size for large enough numbers) population of
  cores that follow the Salpeter distribution over the same mass
  range. We carry out a Kolmogorov-Smirnov (KS) test with synthetic populations of cores by 
  generating the corresponding number of random masses uniformly distributed in each mass bin and repeat for 50 times. 
  We find that the median $p$ value, which indicates the probability that these distributions are consistent with the same
  parent distribution, is $\lesssim 10^{-4}$. Thus we conclude that our estimated
  CMF in IRDC environments is top heavy compared to Salpeter. Such a
  conclusion has also recently been reported in the more evolved
  ``mini-starburst'' W43 region by Motte et al. (2018).

\subsection{Comparison to G286}

\begin{figure*}[htbp]
\centering{\includegraphics[width=0.8\textwidth]{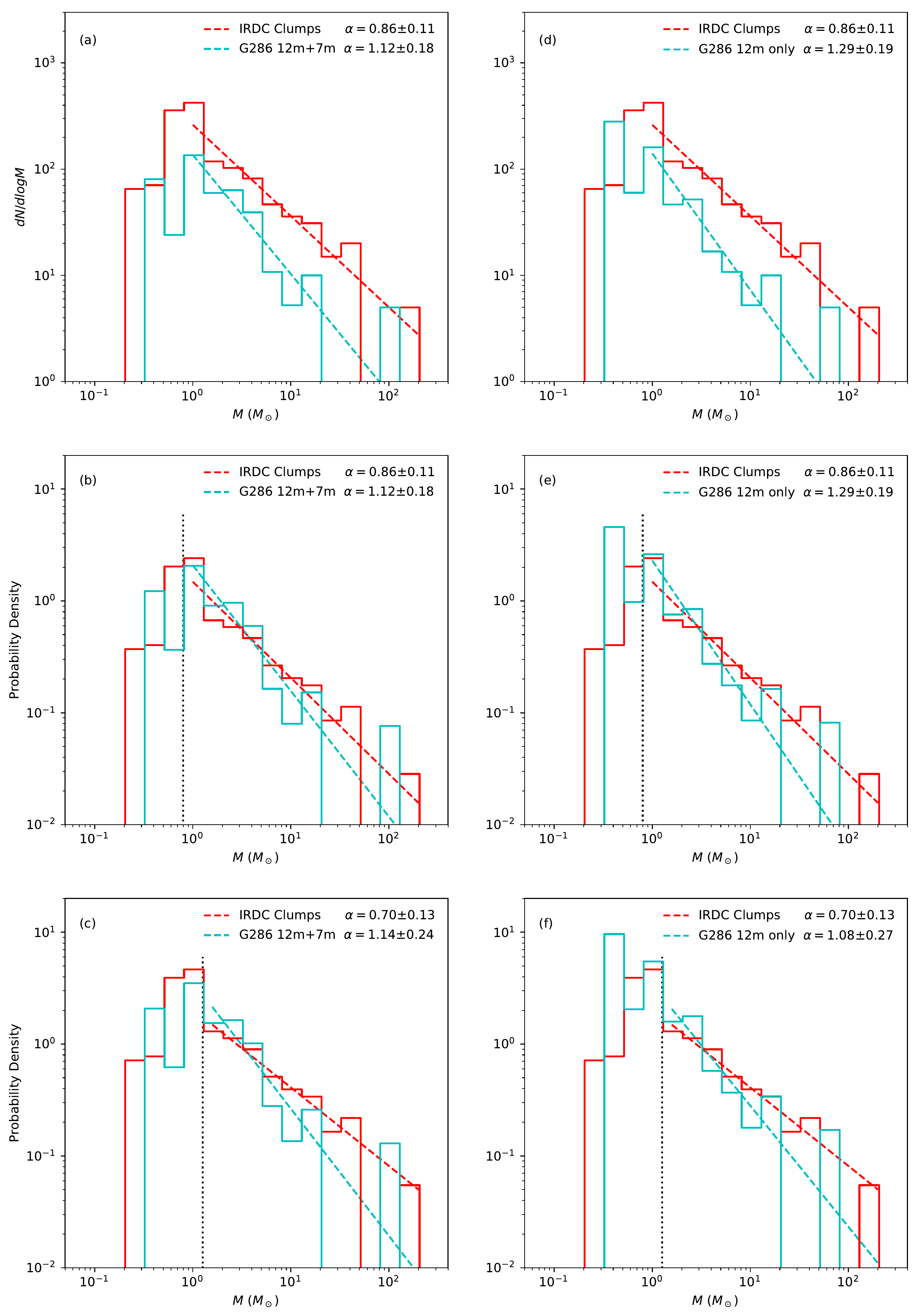}}
\caption{
{\it (a) Top:} Flux and number-corrected ``true'' dendrogram-derived
CMF of IRDC clumps (red histogram) compared with the same CMF derived
from G286 in Paper I (cyan histogram). The simple power law fits to $M\geq0.79\:M_\odot$ are also indicated.
{\it (b) Middle:} As (a), but now showing CMFs normalized by the
number of cores estimated, after completeness corrections, to have $M\geq0.79\:M_\odot$, i.e., 176 cores in the IRDC Clumps and 66 cores in G286 after completeness corrections. This mass threshold is indicated by the vertical black dotted line.
{\it (c) Bottom:} As (a), but now showing CMFs normalized by the
number of cores with $M\geq1.26\:M_\odot$, i.e., 91 cores in the IRDC Clumps and 39 cores in G286 after completeness corrections. This mass threshold is indicated by the vertical black dotted line. Simple power law fits for this mass range are also shown.
{\it (d) Top:} As (a), but comparing to results from G286 12m-only data. 
{\it (e) Middle:} As (b), but comparing to results from G286 12m-only data (61 cores in G286 after completeness corrections are used for the normalization).
{\it (f) Bottom:} As (c), but comparing to results from G286 12m-only data (29 cores in G286 after completeness corrections are used for the normalization).
}\label{fig:G286}
\end{figure*}

Here we present a detailed comparison of our fiducial
dendrogram-derived CMF in IRDC clumps with that measured in the more
evolved G286 protocluster in Paper I. We have already noted and
summarize again that there are some unavoidable differences in our
observational data and analysis methods compared to Paper I. In
addition to the primary beam effect mentioned in \S\ref{S:methods_ci}, our
observations do not include the 7m array and so lack sensitivity to
larger-scale structures. Also, we compile a CMF from observations of
multiple clouds that are at a range of distances, whereas Paper I
studied a single protocluster, G286, at a single distance of
2.5~kpc. We will compare to the results of the 1.5\arcsec resolution
analysis of Paper I, since, as discussed below, this is a better match
to our observations of typically more distance IRDCs at $\sim$1\arcsec
resolution.

Figure~\ref{fig:G286}a shows shows the dendrogram-derived
flux and number-corrected, i.e., ``true'' CMFs from the IRDC clumps
and G286 together. Figure~\ref{fig:G286}b shows these same CMFs, but
now normalized by the number of cores they contain in the $1\:M_\odot$
mass bin and greater, i.e.,
$M\geq0.79\:M_\odot$. Figure~\ref{fig:G286}c shows the CMFs normalized
by the number of cores they have with $M\geq1.26\:M_\odot$, i.e., in
case the $1\:M_\odot$ mass bin is adversely affected by systematic
errors, especially from IRDC D. This panel also displays the power law
indices that result from simple fitting over this slightly higher mass range.

The potential systematic difference resulting from the lack of 7-m array
data for the IRDC clumps needs to be considered. Paper I
found that the CMF derived without 7-m array data in G286 is steeper
by about 0.1. Accounting for this effect thus
may accentuate the difference between the IRDC clump and G286 CMFs. We proceed to re-analyze the G286 data but now excluding the 7m-array data, which gives the fairest comparison with our IRDC clump observations. These results are shown in Figure~\ref{fig:G286} d, e and f.

We carry out a KS test of the high-mass end CMFs to
see if the distributions identified in IRDC clumps and in G286 (with 12m only data) are
consistent with being drawn from the same parent distribution. For the
case of CMFs in the range $M\geq0.79\:M_\odot$, the resulting $p$
value is 0.42. For the distributions in
the range $M\geq1.26\:M_\odot$, the KS test yields $p=0.23$.
Thus these results indicate that the distributions are possibly consistent with one another, in spite of the apparent differences in their power law indices. If we were to boost the number of cores by a factor of 5 and keeping the same distributions, then the $p$ values would become smaller to the point that they would be inconsistent with one another. This test indicates that such an increase in sample size is needed to be able to distinguish between CMFs that have a difference in $\alpha$ of about 0.4.


One potential systematic effect resulting from differences between
the observations is that G286 is at a single distance of $d=2.5\:$kpc
and was observed with a resolution of about 1.5\arcsec and with a
noise level of 0.5~mJy~beam$^{-1}$, while the IRDCs, are observed with
a resolution of $\sim1\arcsec$ and noise level of
$\sim0.2$~mJy~beam$^{-1}$. Paper I also presented results for
1.0\arcsec resolution and with a noise level of 0.45~mJy~beam$^{-1}$,
which yields $\alpha=1.24\pm0.17$ for $M\geq0.79\:M_\odot$, however,
we have decided to focus on the lower resolution resolution results,
given that the IRDCs span a range of distances from 2.4 to 5.7~kpc,
but with IRDC C at 5~kpc and IRDC D at 5.7~kpc contributing a large
fraction of the sample so that the average distance of the IRDC cores
is 4.4~kpc. Thus in the end the effective linear resolutions are
similar (within about 15\%) for the average IRDC core and that
achieved in G286. Overall the mass sensitivities are also quite
similar between the two observations and the completeness correction
factors are relatively modest, at least for $M\geq 1.26\:M_\odot$.

\section{Discussion and Conclusions}\label{S:conclusions}

We have measured the CMF in a sample of about 30 IRDC clumps,
including accounting for flux and number recovery incompleteness
factors. With simple fitting, we derived high-end power law indices of
$\alpha \simeq 0.86 \pm 0.11$ for $M\geq 0.79\:M_\odot$ and
$\alpha \simeq 0.70\pm0.13$ for $M\geq 1.26\:M_\odot$. An MLE analysis
yielded similar values. These results indicate a CMF that is top heavy compared the standard Salpeter distribution with $\alpha=1.35$.

To reduce the potential effects of systematic uncertainties, we have
compared the above results to the CMF derived with similar methods in
the more evolved protocluster G286 (Paper I).  From the considerations
of \S\ref{S:results}, we expect that the most reliable comparison is
for the higher mass range of the CMF, $M\geq 1.26\:M_\odot$, for which
we have found $\alpha=1.08\pm0.27$ for G286 when only the 12m-array
data are analyzed.
These results thus indicate only a hint of a potential variation
in the high-mass end of the CMF between the Galactic environments of
IRDC clumps (i.e., early stage, high pressure centers of
protoclusters) and G286, i.e., a more evolved protocluster that is
sampled more globally, i.e., both central and outer regions. One
  of the main factors limiting our ability to distinguish the
  distributions is the relatively small number of cores in each of the
  samples used in this direct comparison. Increasing the sample by
  about a factor of 5 is expected to enable these distributions to be
  reliably distinguished, if they maintain their currently observed
  forms.

Overall, the values of power law index of the CMF derived in
G286 is similar to that of the Salpeter stellar IMF, i.e.,
$\alpha=1.35$, while that in the IRDC clumps is shallower,
indicating relatively more massive cores are present. This may
indicate that massive stars are more likely to form in high mass
surface density, high pressure regions of IRDCs. Such a difference in the CMF
and resulting IMF could potentially be caused by a number of different
physical properties of the gas that vary systematically between the
regions. On the one hand, the higher density, higher pressure regions
of IRDC clumps is expected to lead to a smaller Bonnor-Ebert mass,
which would also take a value $\ll1\:M_\odot$ (see, e.g., McKee \& Tan
2003). The fact that we see evidence for a more top-heavy CMF
indicates that thermal pressure is not the main factor resisting
gravity in setting core masses in these environments, which would then
indicate that some combination of increased turbulence and/or magnetic
field support is present in IRDC clumps.

Note that IRDC clumps are cold regions, so that extra thermal heating
of the ambient environment from radiative feedback from surrounding
lower-mass stars, as proposed in the model of Krumholz \& McKee
(2008), is not expected to be greater here compared to more evolved
stages as represented by G286. However, localized heating of the core
from the protostar itself is expected to be higher in higher mass
surface density environments, if powered mostly by accretion (Zhang
\& Tan 2015). At the moment we do not have any direct indication if
the localized temperatures of cores are higher in the IRDC clumps
compared to G286. Note, if localized IRDC core temperatures were
systematically higher than in G286, then we would have overestimated
their masses. If this effect is greater for the more luminous mm cores
and is systematically greater in the IRDC sample compared to G286,
then this would make their intrinsic CMFs more similar. Such
considerations highlight additional potential systematic effects due to
temperature or dust opacity variations that need to be treated as
caveats to our results, and indeed all results of CMFs derived from mm dust
emission when individual core temperature and opacity data are not
available.

Comparing with previous studies in IRDCs, our relatively flat
high-mass end power law index is consistent with the results of Ragan
et al. (2009), although they probed a different mass range of 30 to
3,000$\:M_\odot$ and used different methods, i.e., MIR extinction,
which is subject to a variety of systematic uncertainties (Butler \&
Tan 2012), including foreground corrections that effect lower column
density regions and ``saturation'' effects at high optical depths
causing the mass in high column density regions to be underestimated.
Zhang et al. (2015) also found a relative lack of lower-mass cores
compared to the Salpeter (1955) distribution, but their sample size
was relatively small (only 38 cores selected in a single small,
$\sim0.5$~pc region) and they did not carry out completeness
corrections. Still, their results do illustrate the effects of using
higher angular resolution (by about a factor of two, i.e.,
$\sim0.8\arcsec$), better sensitivity (by about a factor of three,
i.e., $1\sigma$ rms of 75$\: \rm \mu$Jy), but with more limited
sensitivity to larger scale structures (given a more extended
configuration of {\it ALMA} was employed) compared to our current
study. The 5 cores we identify in the C2 clump are further decomposed
into 34 cores by Zhang et al. (2015), i.e., the bulk of their sample,
in their analysis of core identification, which is based on the
dendrogram method, but also supplemented by dendrogram-guided Gaussian
fitting of additional structures. On the other hand, Ohashi et
al. (2016) found a steeper power law index for the pre-stellar CMF
derived in their study (28 cores in IRDC G14.225-0.506 found by 3~mm
continuum emission), although the uncertainty in their result is large
($\alpha \simeq 1.6\pm0.7$) and, again, their methods differ from
ours, especially the lack of completeness corrections for flux and
number. Motte et al. (2018) have recently studied the 1.3~mm dust
continuum derived CMF in the W43-MM1 ``mini-starburst region'',
finding $\alpha=0.90\pm0.06$ for $M>1.6\:M_\odot$, based on a sample
of 105 cores. We note they used different methods of core
identification, i.e., the {\it getsources} algorithm (Men'shchikov et
al. 2012), but also carried out a visual inspection step of removing
cores that were ``too extended, or whose ellipticity is too large to
correspond to cores, or that are not centrally-peaked'', so a direct
comparison with our results is not as meaningful as our comparison to
the G286 protocluster.

In summary, we see that quantitative direct comparison of our results
with these previous studies is not particularly useful given the
differences in the data and methods used to identify cores and
estimate CMFs. We thus emphasize that, in addition to finding a more
top heavy CMF compared to the Salpeter distribution, our main result
for a hint of a potential variation in the CMF in different
environments is based on the comparison with our Paper I study of
G286, which used more similar data and methods.

Future progress in this field can take several directions. First, as
discussed above, much larger samples of cores in these types of
environments are needed. Second, a wider range of Galactic
environments need to be probed. Third, the CMF should be probed to
lower masses to better determine the location of any peak. This will
require higher sensitivity and higher angular resolution
observations. Such observations will also likely change the shape of
the high-mass end of the CMF by sometimes breaking up more massive
``cores'' into smaller units. Fourth, better constraints on potential
systematic effects related to mass determination from mm continuum
flux are needed, especially by individual temperature measurements of
the cores. Fifth, the evolutionary stage of the cores should be
determined, i.e., protostellar mass to core envelope mass, including
determining if cores are pre-stellar, i.e., via astrochemical
indicators or via an absence of outflow indicators or concentrated
continuum emission. Such information is needed to better determine how
the CMF and IMF are actually established in protocluster environments,
as discussed by Offner et al. (2014).

\acknowledgments We thank Paola Caselli and Francesco Fontani for helpful comments on the manuscript. We thank Adam Ginsburg for helpful discussion and suggestions. J.C.T. acknowledges NSF grants AST1411527. This paper makes use of the following ALMA data: ADS/JAO.ALMA\#2013.1.00806.S. ALMA is a partnership of ESO (representing its member states), NSF (USA) and NINS (Japan), together with NRC (Canada), NSC and ASIAA (Taiwan), and KASI (Republic of Korea), in cooperation with the Republic of Chile. The Joint ALMA Observatory is operated by ESO, AUI/NRAO, and NAOJ. The National Radio Astronomy Observatory is a facility of the National Science Foundation operated under cooperative agreement by Associated Universities, Inc.


\listofchanges

\end{document}